\documentclass[%
 reprint,
 superscriptaddress,
 amsmath,
 amssymb,
 aps,
 longbibliography
]{revtex4-2}
\usepackage{graphicx}
\usepackage{dcolumn}
\usepackage{bm}
\usepackage{graphicx, subfig}
\usepackage[colorlinks,linkcolor=black,anchorcolor=black,citecolor=black,urlcolor=black]{hyperref}
\usepackage[T1]{fontenc}
\usepackage{enumitem}
\DeclareMathOperator{\tr}{Tr}
\usepackage[justification=raggedright]{caption}
\usepackage{multirow}
\usepackage{tikz}
\usepackage{array}
\usepackage{booktabs}
\usepackage{siunitx}
\usepackage{makecell}
\usepackage{rotating}
\usepackage{colortbl}
\usepackage{upgreek}
\usepackage{xcolor}
\definecolor{lightgray}{gray}{0.8}

\begin{document}

\preprint{APS/123-QED}

\title{Machine Eye for Defects: Machine Learning-Based Solution to Identify and Characterize Topological Defects in Textured Images of Nematic Materials}

\author{Haijie Ren}
\affiliation{%
Department of Physics, The Hong Kong University of Science and Technology, Clear Water Bay, Kowloon, Hong Kong, P. R. China
}%

\author{Weiqiang Wang}
\affiliation{%
Department of Physics, The Hong Kong University of Science and Technology, Clear Water Bay, Kowloon, Hong Kong, P. R. China
}%

\author{Wentao Tang}
\affiliation{%
Department of Physics, The Hong Kong University of Science and Technology, Clear Water Bay, Kowloon, Hong Kong, P. R. China
}%


\author{Rui Zhang}
\email{ruizhang@ust.hk}
\affiliation{%
Department of Physics, The Hong Kong University of Science and Technology, Clear Water Bay, Kowloon, Hong Kong, P. R. China
}%

\date{\today}

\begin{abstract}
Topological defects play a key role in the structures and dynamics of liquid crystals (LCs) and other ordered systems. There is a recent interest in studying defects in different biological systems with distinct textures. However, a robust method to directly recognize defects and extract their structural features from various traditional and nontraditional nematic systems remains challenging to date. Here we present a machine learning solution, termed Machine Eye for Defects (MED), for automated defect analysis in images with diverse nematic textures. MED seamlessly integrates state-of-the-art object detection networks, Segment Anything Model, and vision transformer algorithms with tailored computer vision techniques. We show that MED can accurately identify the positions, winding numbers, and orientations of $\pm 1/2$ defects across distinct cellular contours, sparse vector fields of nematic directors, actin filaments, microtubules, and simulation images of Gay--Berne particles. MED performs faster than conventional defect detection method and can achieve over 90\% accuracy on recognizing $\pm1/2$ defects and their orientations from vector fields and experimental tissue images. We further demonstrate that MED can identify defect types that are not included in the training data, such as giant-core defects and defects with higher winding number. 
Remarkably, MED provides \textcolor{black}{correct structural information about $\pm 1$ defects, i.e., the phase angle for $+1$ defects and the orientation angle for $-1$ defects.} As such, MED stands poised to transform studies of diverse ordered systems by providing automated, rapid, accurate, and insightful defect analysis.
\end{abstract}
\maketitle


\section{\label{Introduction}Introduction}
\textcolor{black}{Topological defects are local regions within an ordered medium where its order is frustrated or changes abruptly. These defects }are ubiquitous and play a salient role across diverse physics disciplines, including liquid crystals (LCs), superfluids, early universe, and optics~\cite{mermin1979topological, de1993physics}. In nematic LCs, the average molecular orientation within LCs allows for the introduction of the director field, indicative of the LC's micro-structure~\cite{ravnik2009landau}. Notably, defects within the director field are characterized by a relatively low scalar order parameter, typically $S < 0.4$~\cite{kleman2003soft}. As the field of LC research has evolved, there is a rapidly growing interest in studying diverse materials systems as nematic LCs, in which topological defects are the focus of research~\cite{zhang2016dynamic, tang2017orientation, zhang2022logic, wang2021interplay, zhang2018interplay}. Examples from biological systems include epithelial cells~\cite{saw2017topological}, progenitor neural cells~\cite{Kawaguchi2017topological}, dense bacteria suspensions~\cite{Li2019datadriven}, and spindles during mitosis~\cite{Jan2014spindle}. Therefore, there is a strong need for a robust method to analyze the director field and defects in distinct systems with nematic texture.


Reflecting upon the past research on LC materials, it is evident that it has benefited from experimental, theoretical, and computational approaches, a traditional scientific paradigm enumerated by Jim Gray, a Turing Award laureate~\cite{JimGray}. However, the landscape of scientific exploration is undergoing a dynamic transformation, driven by notable reductions in computational and data storage costs~\cite{Rudy,raissi2019,raissi2020}. This revolution in the research ecosystem is fostering an emergent shift towards a data exploration paradigm --- a realm where data-driven discovery takes precedence. Central to this paradigm is machine learning (ML), which has shown remarkable progress over the past decade, as evidenced by breakthroughs such as ResNet~\cite{He_2016_CVPR} in Image Recognition and Transformer models~\cite{vaswani2017attention} across various language tasks. Large language Models, such as GPT-4~\cite{openai2023gpt4}, are built upon the Transformer model, currently attracting significant attention, implying a promising future for ML in physics.

The adoption of this emerging data-driven methodology has begun to gain traction in LC research~\cite{Ingo1,Ingo2,Ingo3,Eric2022,colen2021machine,PInematic2023,minor2020end,chowdhury,Ribeiro2019,Ribeiro2022}. For instance, Colen et al. employed Convolutional Neural Networks (CNN) and Long Short Term Memory (LSTM) for predicting the activity and elastic constant of active nematics, demonstrating the potential for forecasting the chaotic-like dynamics of active defects~\cite{colen2021machine}. Similarly, Golden et al. utilized a symbolic regression-based method to develop a mathematical model of active nematics from experimental data, remarkably recovering the Leslie--Ericksen model~\cite{PInematic2023}. In these efforts, however, topological defects that are imperative to active nematics phenomena are either implicitly included in the model or explicitly excluded from the data.

\begin{figure*}[htbp]
\includegraphics[width=1\textwidth]{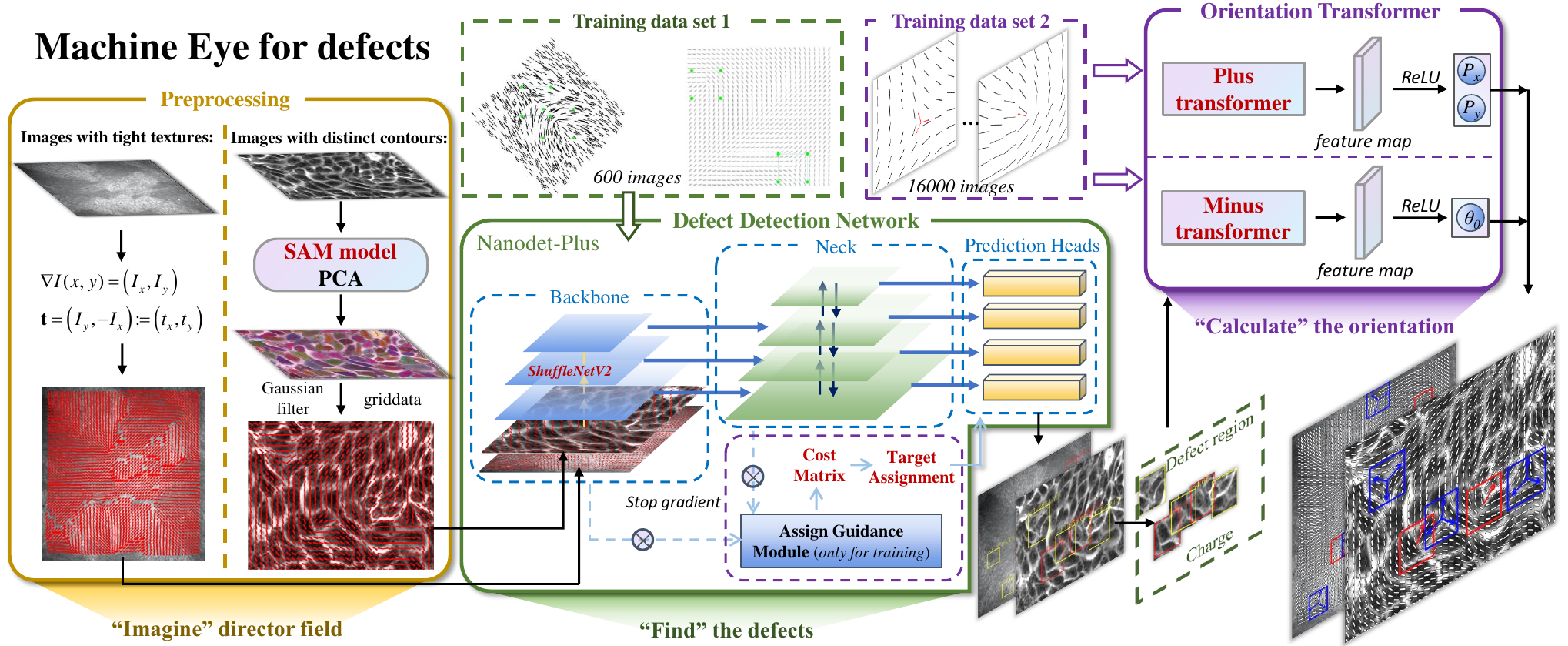}
\caption{\label{MEDsystem} \textbf{The Machine Eye for Defects (MED) System: A Schematic Overview.} The preprocessing modules are tasked with the extraction of an equivalent-sized director field from the provided image, akin to how we would ``imagine'' director field. Following this, Defect Detection Network is employed to ``find'' the defects' position and corresponding topological charge. Finally, the Orientation Transformer is assigned to predict or ``calculate'' 
\textcolor{black}{the orientations of all of the detected defects}.}
\end{figure*}


The challenge of directly recognizing defects has just begun to be addressed using ML techniques. \textcolor{black}{Ronhovde et al.'s early work implemented unsupervised ML method for the systematic detection and analysis of structures in various complex physical systems, with a specific emphasis on identifying defects in crystals and discerning patterns in amorphous materials like glasses~\cite{Ronhovde2011EPJE,ronhovde2012detection}. Recent advancements in ML have revolutionized the detection of defects in LCs.} Walters et al. trained Deep Neural Networks (DNN) and LSTM using the $(x,y,\theta)$ coordinates for a two-dimensional (2D) director field containing $-1/2$ or $-1$ defects, and achieved the classification of defect types in testing director fields~\cite{walters2019machine}. Minor et al. trained You Only Look Once (YOLO) v2~\cite{yolov2} with simulated cross-polarisation images of defects, thereby enabling pinpointing defect locations in cross-polarised images~\cite{minor2020end}. 
Building upon this, Chowdhury et al. \textcolor{black}{enhanced their approach by training YOLOv5~\cite{YOLOv5} with experimental cross-polarized images to locate $\pm1$ defects in a smectic LC film. They cross-correlated defect core regions with angled synthetic templates to reveal orientation dynamics of surrounding brush-like structures. These dynamics are then processed through a binary classification network to predict the topological charges of the defects~\cite{chowdhury}.} \textcolor{black}{Li et al.~\cite{li2023machine} developed a method where they trained ResNet~\cite{He_2016_CVPR} blocks using experimental microtubule images and the corresponding orientation fields obtained via PolScope. This approach enables direct prediction of the orientation field from experimental images. Subsequently, they calculated the winding numbers of the defects from the director field and further obtained defect orientations using additional ResNet blocks.}

Despite the above recent advances in using ML to identify defects in LCs, there are two groups of open questions remained to be addressed. 
Scientifically, we are questioning the generality or extensibility of ML algorithms. For example, can an ML algorithm identify defects that are beyond the training data? And beyond the positions of the defects, can an ML algorithm identify their structural information \textcolor{black}{from images}, such as their winding number, orientation, \textcolor{black}{or phase angle}?
Practically, existing ML attempts often-times rely on a specific type of experimental images or nematic textures. Here we ask: is there a robust ML package that can work with arbitrary images of LC textures? And whether data-driven methods can outperform the physics-based traditional defect tracking method?

To answer these open questions, in this work, we harness cutting-edge object detection algorithms~\cite{=nanodet}, Segment Anything Model (SAM)~\cite{kirillov2023segment}, Vision Transformer~\cite{vit2020image,wu2020visual,deng2009imagenet} and computer vision (CV) techniques to analyze the textured images of nematic materials. Our technique, termed Machine Eye for Defects (MED), transcends simple defect detection methods by its capability to identify the positions, winding numbers, and orientations of defects from images with distinct nematic textures. Moreover, we show that MED can comprehend defects that are not included in the training data, demonstrating its potential to be generalized. The specifics of MED and its application to images of different nematic systems for defect detection will be discussed in details in the following sections.

\section{\label{MED system}Method}

MED consists of three key modules: a preprocessing module that employs the Segment Anything Model (SAM) and computer vision (CV) techniques, a defect detection network based on the object detection model, i.e., Nanodet-plus, and a defect orientation prediction module based on the Vision Transformer paradigm. The intricate architecture of the MED framework is visually 
\textcolor{black}{depicted} in Fig.~\ref{MEDsystem}. In order to ensure the accessibility and ease of use for MED, we performed all the 
\textcolor{black}{training and testing} on a laptop equipped with an NVIDIA RTX3060 GPU.

\subsection{\label{Preprocessing}Preprocessing module}
Our preprocessing module adeptly addresses images with distinct textures, such as discrete contours (e.g., epithelial cells or vector fields of directors) and tight textures (e.g., cytoskeletal filaments).


For images with discrete contours, a distinct segment in the preprocessing module is employed. Traditional CV methods, which often necessitate manual parameter tuning across images, are inefficient. Here, we integrate the SAM method~\cite{kirillov2023segment}, which excels in contour delineation without the need for additional training. This enables effective segmentation of cellular structures into distinct contours during cellular image analysis. A function for quantifying segmentation overlap is also included to bolster the model's robustness (Appendix~\ref{Technical detail}). Subsequently, Principal Component Analysis (PCA)~\cite{scikit-learn,abdi2010PCA} is 
\textcolor{black}{used} to calculate the dominant orientation of each segmented contour, which is used for further interpolation and refinement of the director field (Appendix~\ref{SAM detail} and~\ref{interpolate detail}).

In the context of tight-textured images such as cytoskeletal filaments, the algorithm is based on the concept that the intensity gradient is orthogonal to the average orientation of the filaments~\cite{boudaoud2014fibriltool}. For high-quality and noise-free images, e.g., Fig.~\ref{Tight textures}(a), the intensity gradient at each pixel directly gives the local director field~\cite{sciortino2023polarity}. However, when noise or blurriness is present in the image, such as Fig.~\ref{Tight textures} (c) and (d), this straightforward director field extraction becomes error-prone. To counteract this noise susceptibility, we extract the director field from local areas using PCA method instead of from individual pixels. More details can be found in Appendix~\ref{interpolate detail}. 


\subsection{Defect detection network}
In various applications, ranging from autonomous driving~\cite{2022yuanKeypoint} to facial recognition and clustering tasks~\cite{Schroff_2015_CVPR}, object detection algorithms serve as a critical component. It is worth noting that the concept of ``object'' is not limited to conventional entities but can also encompass topological defects. 
For instance, Minor et al.~\cite{minor2020end} and Chowdhury et al.~\cite{chowdhury} employed YOLO \textcolor{black}{in their research}, a prominent object detection framework. These detection networks predominantly utilize Convolutional Neural Networks (CNNs) and feature classification mechanisms based on annotated labels. Our work aims to extend this paradigm by not only identifying defect positions but also characterizing their winding number and orientation. After evaluation, Nanodet-Plus~\cite{=nanodet} emerged as the most suitable Keypoint-based object detection algorithm, leading us to construct the defect detection network. Further details regarding the selection criteria and methodological details for object detection algorithms are provided in the Appendix~\ref{Detectdetail}.

The defect detection network was trained on a dataset comprising approximately 600 labeled images of sparse director field, each paired with specific defect labels. Each image possesses one $+1/2$ defect and one $-1/2$ defect, constructed from the hybrid lattice Boltzmann simulations (Appendix~\ref{Train data}). These images typify the `Training data set 1' in Fig.~\ref{MEDsystem}. The computational cost for this training process was approximately 7 hours. It is essential to precisely position the defect core at the label box's center during labeling, allowing it to serve as the keypoint of algorithms. Despite the uniformity of director field images and limited training sample size, our defect detection network exhibited commendable efficacy in subsequent intricate recognition tasks.

\subsection{Orientation Transformer}

The transformer models~\cite{vaswani2017attention} have become noteworthy for their high performance in diverse language tasks, solidifying their role as the standard for natural language processing. They have been extended to computer vision applications, notably through the Vision Transformer (ViT)~\cite{vit2020image}. We integrated ViT into our Orientation Transformer for effective feature extraction and accurate prediction. A feature map as an output from the ViT is fed into another neural network through a Rectified Linear Unit (ReLU) activation function (Fig.~\ref{MEDsystem}). 

Two specialized transformers are 
\textcolor{black}{used} to predict defect orientations: `Plus Transformer' outputs a coordinate pair $(P_x, P_y)$, representing the components of the orientation vector ${\bf P}$ associated with the $+1/2$ defect; and `Minus Transformer' produces $\theta_0$ for the $-1/2$ topological defect, which is an angular phase variable capable of characterizing the defect's orientation (Appendix~\ref{Train data}). When a defect is detected, defect detection network sends to the Orientation Transformer the image of the defect region and the corresponding topological charge, prompting activation of the relevant transformer for orientation prediction. The Orientation Transformer structure details can be seen in Appendix~\ref{Technical detail}. 

To keep the training cost manageable, we initially employed the pre-trained model `vit-base-patch16-224-in21k'~\cite{wu2020visual,deng2009imagenet} with an equivalent architecture for feature extraction. We fine-tuned the ViT model's gradients and trained the Orientation Transformer with our `Train dataset 2', including 16,050 $\pm 1/2$ defect images and the corresponding $(P_x, P_y)$ and $\theta_0$. The computational cost for training the orientation transformer was approximately 16 hours. Further details on methodological details and training data can be found in Appendices~\ref{detailtransformer} and~\ref{Train data}.

\begin{figure*}[htbp]
\includegraphics[width=1\textwidth]{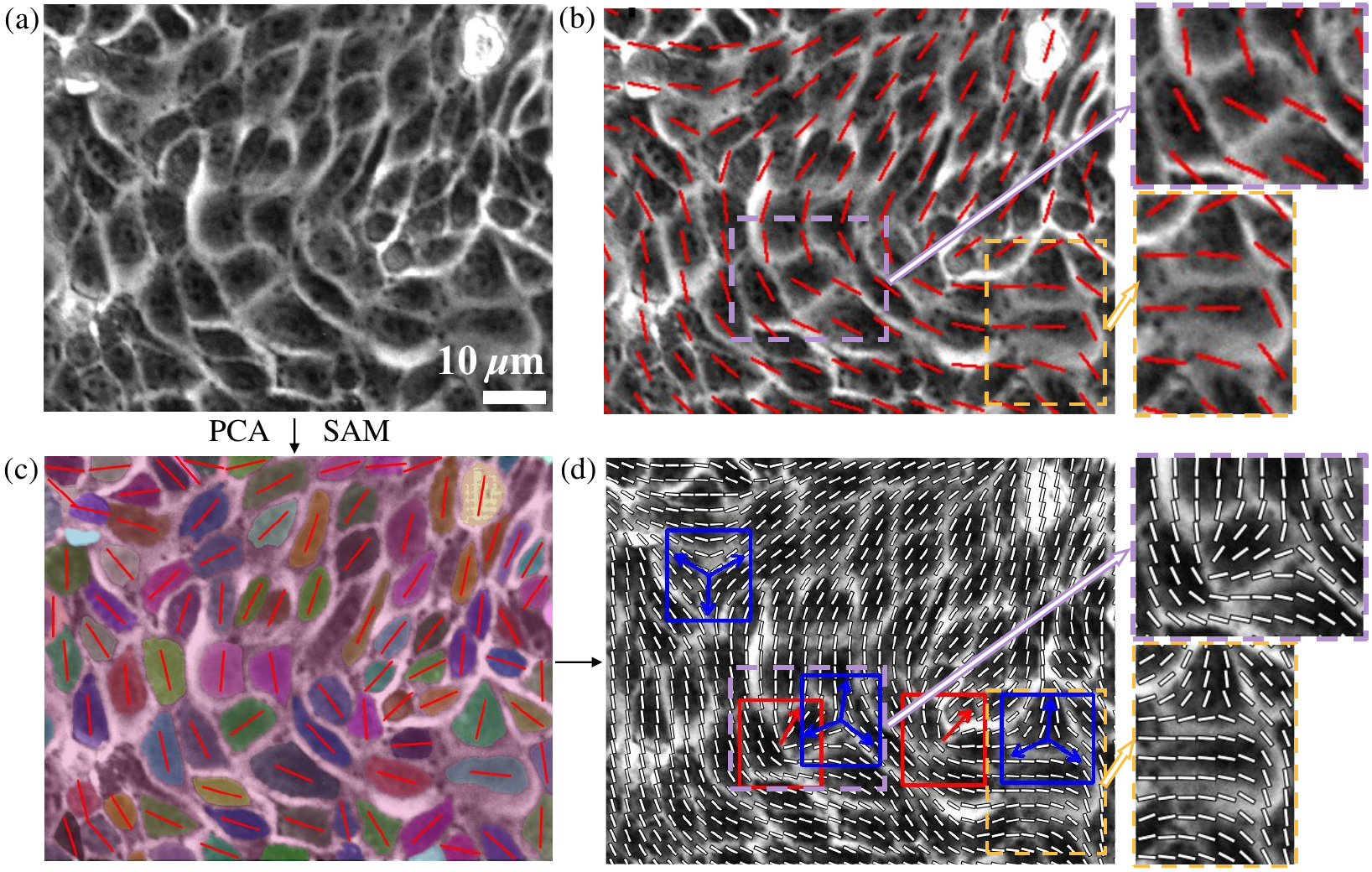}
\caption{\label{Cells} \textbf{Topological defect identification in Epithelial Cells.} (a) Wild type epithelial MDCK cell \textcolor{black}{micrograph} from Ref.~\cite{saw2017topological}. (b) Nematic director field corresponding to (a) presented in Ref.~\cite{saw2017topological}. (c) The Preprocessing module utilizes SAM~\cite{kirillov2023segment} to segment cellular regions and employs PCA to identify each cell's principal axis. (d) MED identifies $\pm 1/2$ topological defect positions, winding numbers, and orientations, represented by arrows or tri-arrows in square boxes, with red denoting $+1/2$ defects and blue denoting $-1/2$ defects. 
\textcolor{black}{The zoomed-in boxes in (b) and (d) show a comparison of the director fields extracted by the two methods.} Scale bar: $10~\upmu$m.}
\end{figure*}

\section{Results}
\subsection{Robustness of MED}

We first consider images of epithelial cells~\cite{saw2017topological} (Fig.~\ref{Cells}(a), (b)), which serve as quintessential examples with discrete contours.
By implementing the SAM model~\cite{kirillov2023segment}, we first extract the local orientations of the cells (Fig.~\ref{Cells}(c)), based on which we generate a smooth director field (Fig.~\ref{Cells}(d)). 
The extracted director field image is subsequently processed by the trained defect detection network to identify and characterize defects within the image. $+1/2$ and $-1/2$ defects are marked with red and blue boxes, respectively. The processing of each defect region through the trained Orientation Transformer allows for the acquisition of the corresponding defect orientations, represented by arrows of corresponding colors (Fig.~\ref{Cells}(d)).
By comparing the director field and defects provided by MED to those reported in literature~\cite{saw2017topological}, we find that MED follows 
\textcolor{black}{long axis of the cells} more faithfully, 
\textcolor{black}{yielding a more detailed analysis of the tissue texture.} In what follows, we discuss more examples showcasing the validity and accuracy of MED.

\begin{figure*}[htbp]
\includegraphics[width=1\textwidth]{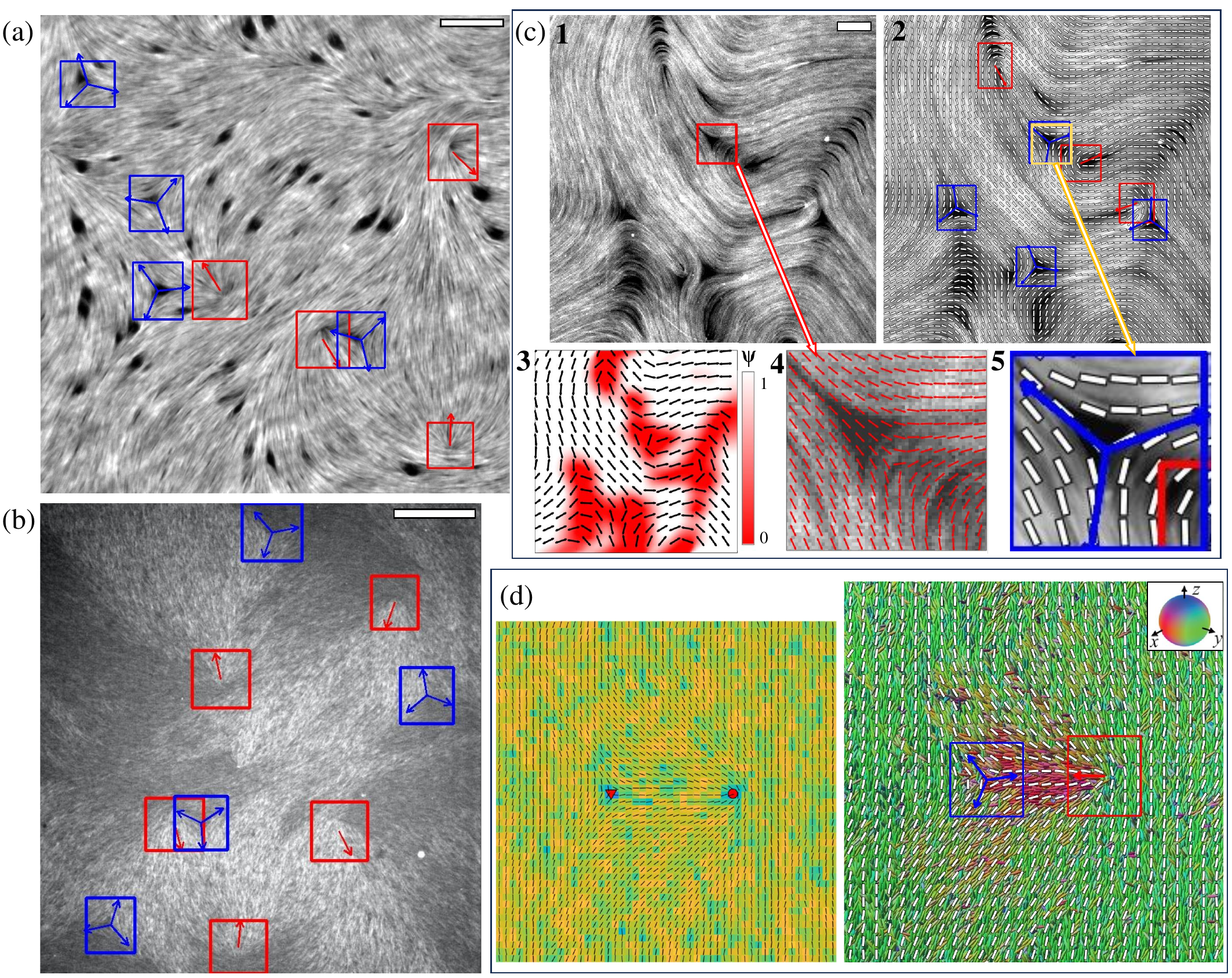}
\caption{\label{Tight textures} \textbf{Defect identification from nematic images with tight textures.} (a) Actin-based nematic images from Ref.~\cite{sciortino2023polarity} marked with MED predicted defects; scale bar: $10~\upmu$m. (b) Actin-based nematic images from Ref.~\cite{zhang2018interplay} marked with MED predicted defects; scale bar: $30~\upmu$m. 
\textcolor{black}{In (c), 1: A snapshot of a microtubule-based active nematic from Ref.~\cite{PInematic2023}; 2: the same image overlaid with MED predicted director field; scale bar: $50~\upmu $m; 3: the director field corresponding to 1 presented in Ref.~\cite{PInematic2023}; 4,5:  the difference in analyzing the director field around a void $-1/2$ defect between the original work (4) and the MED result (5). (d) Right: The background displays an MD simulation snapshot of a pair of $\pm1/2$ defects in a nematic comprising Gay--Berne particles~\cite{huang2023structures}, the inset shows the colormap representing particle orientation; the overlaid section is director field and defects analyzed by MED. Left: the corresponding director field computed by physics-based method directly from simulation datafiles.}}
\end{figure*}

We next study tight nematic textures. We challenge MED by applying it to four different nematic systems as summarized in Fig.~\ref{Tight textures}. For a nematic comprising short actin filaments~\cite{sciortino2023polarity}, MED can correctly identify the positions and orientations of $\pm1/2$ defects despite the presence of void regions of tactoids (Fig.~\ref{Tight textures}(a)). In another type of actin-based nematic image complicated by polarized light illuminations~\cite{zhang2018interplay}, MED does equally well in identifying defects in spite of the spatially variant lightness of the texture (Fig.~\ref{Tight textures}(b)). 
The prediction quality of Fig.~\ref{Tight textures}(a) and (b) demonstrates the MED's robust efficacy in predicting nematic images with different textures~\cite{sciortino2023polarity,zhang2018interplay}. 
\textcolor{black}{In microtubule- or actin-based active nematics, certain areas, such as $-1/2$ defect cores, often appear as void (dark) regions with a depletion of cytoskeletal polymers. For an image containing such void defect regions~\cite{PInematic2023},} MED performs better than existing methods in terms of correctly identifying, for example, $-1/2$ defect positions (Fig.~\ref{Tight textures}(c)). Note that both void tactoids and void defects are not included in our training data.
Beyond experimental images of passive and active nematics, MED can also be extended to analyze simulation images of nematic materials. By analyzing the snapshots from a simulation of a coarse-grained molecular model, i.e., Gay--Berne particle system~\cite{huang2023structures}, MED gives rise to reasonable director fields and defect information (Fig.~\ref{Tight textures}d), laying the groundwork for further explorations into new domains of liquid crystal science and beyond.


\subsection{\label{Accuracy} Predictive Accuracy and Efficiency}

\begin{figure*}[htbp]
\includegraphics[width=1\textwidth]{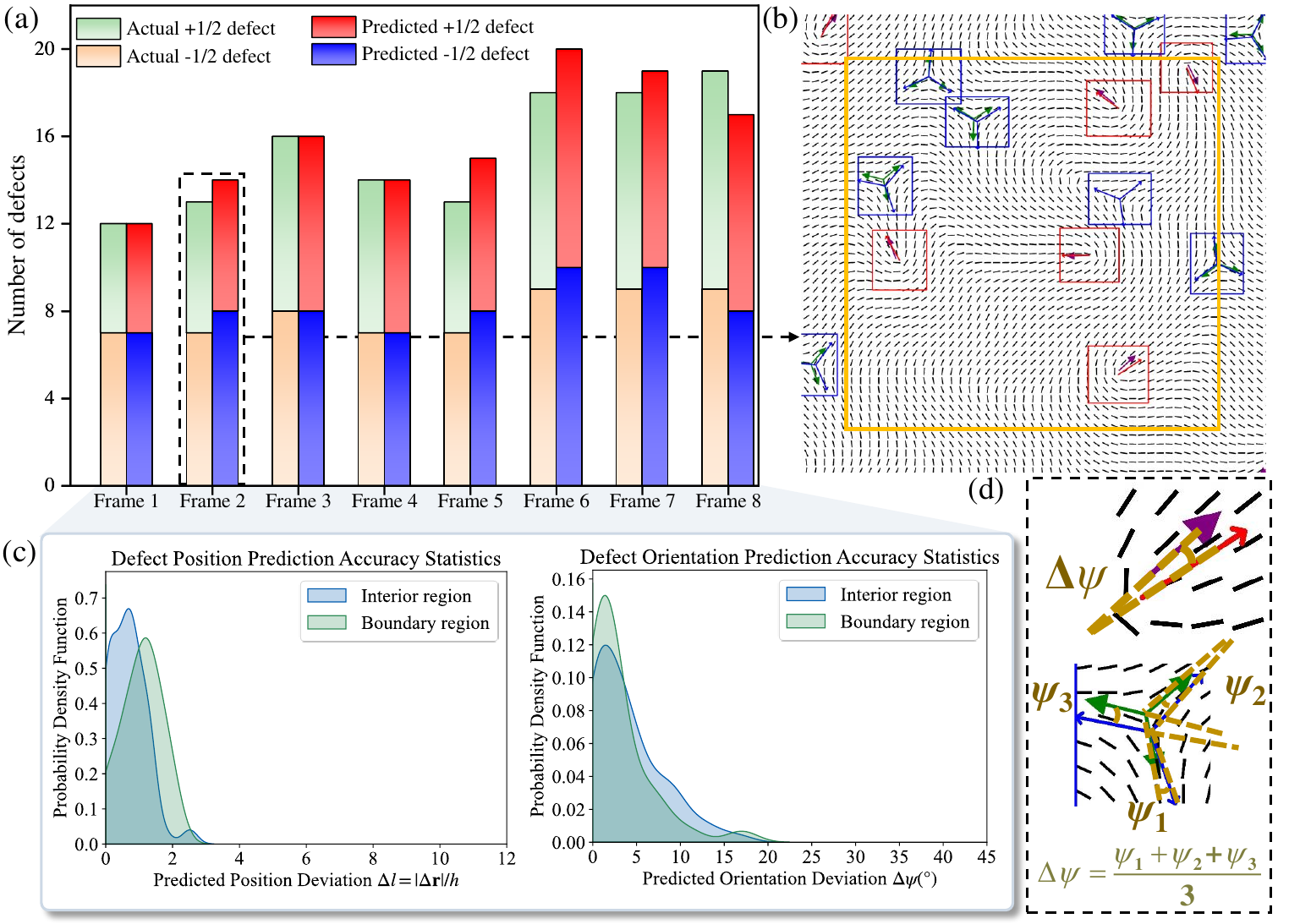}
\caption{\label{Turbulence} \textbf{Accuracy analysis of MED prediction for SI Movie 1.} (a) Comparison between topological defects number as predicted by MED and those from the ground truth. (b) A snapshot of a nematic director field with predicted defects with red and blue representing MED results and purple and green denoting the ground truth. (c) Histogram of the position deviation \textcolor{black}{$\Delta l$} (left) and orientation deviation $\Delta \psi$ (right) from the ground truth for 120 defects using MED. (d) The definitions of $\Delta \psi$ for $\pm 1/2$ defects. \textcolor{black}{For a $+1/2$ defect, the angle $\Delta \psi$ measures the deviation between the predicted orientation (red arrow) and the ground truth (purple arrow). For a $-1/2$ defect, $\Delta \psi_1$, $\Delta \psi_2$, and $\Delta \psi_3$ represent the angle differences between three pairs of predicted (blue arrows) and actual (green arrows) orientation.}}
\end{figure*}

To rigorously assess the predictive accuracy of MED, we have performed two calculations. The first calculation is based on sparse vector field images generated from hybrid lattice Boltzmann method (LBM) simulations (Appendix~\ref{LBM} and Ref.~\cite{zhang2016lattice}). SI Movie 1 shows defects' positions, winding numbers, and orientations predicted by the MED alongside with the corresponding ground truth. The simulations included 123 defects (60 $+1/2$ defects and 63 $-1/2$ defects), with the MED accurately identifying 120 of them, achieving a remarkable accuracy of 97.56\%. Only three defects were unaccounted for, two of which were on the verge of annihilation.

\begin{table*}[htbp]
    \centering
    \caption{\textcolor{black}{Computational efficiency comparison of} traditional physics-based calculation and MED}
    \begin{tikzpicture}
        \draw[->, thick] (0,0) -- (10,0);
        \foreach \x/\y in {0/Step 1, 2.5/Step 2, 5/Step 3, 7.5/Step 4} {
            \draw (\x,-0.1) -- (\x,0.1);
            \node[below] at (\x,-0.2) {\y};
        }
        \foreach \x/\y in {0/Director field, 2.5/Position, 5/Winding number, 7.5/Orientation} {
            \node[above] at (\x,0.2) {\y};
        }
    \end{tikzpicture}
    \begin{tabular}{clcccc}
        \toprule
        \multirow{2}{*}{\centering\thead{Case}} & \multirow{2}{*}{\centering\thead{Method}} & \multicolumn{4}{c}{\thead{Steps}} \\
        \cmidrule(lr){3-6}
        & & Step 1 & Step 2 & Step 3 & Step 4 \\
        \midrule
        \multirow{3}{*}{MD simulation of Gay--Berne particles} & MED & \SI{8.7}{s} & \multicolumn{3}{>{\columncolor{lightgray}}c}{\SI{4.4}{s}} \\
        \addlinespace[0.2em]
        & Physics-based in Matlab & \multicolumn{3}{>{\columncolor{lightgray}}c}{\SI{83}{s}} & $-$ \\
        \addlinespace[0.2em]
        & Physics-based in Python & \multicolumn{2}{>{\columncolor{lightgray}}c}{\SI{11.21}{s}} & $-$ & $-$ \\
        \midrule
        \multirow{2}{*}{LBM simulation of Active turbulence} & MED & $-$ & \multicolumn{3}{>{\columncolor{lightgray}}c}{\SI{39.041}{s}} \\
        \addlinespace[0.2em]
        & Physics-based in C language & $-$ & \multicolumn{3}{>{\columncolor{lightgray}}c}{\SI{37.0}{s}} \\
        \bottomrule
    \end{tabular}
    \label{Efficiency}
\end{table*}

Additionally, eight spurious defects were detected, predominantly located at the boundary of the simulation domain. By introducing false positive rate (FPR) defined as
\begin{equation}
\text{FPR} = \frac{\text{FP}}{\text{TN} + \text{FP}},
\end{equation}
with FP the number of False Positives and TN the number of True Negatives, we find that the resultant FPR is 6.25\%.
To better elucidate the prediction accuracy of MED, the simulation domain was partitioned into boundary and interior regions. The interior region is demarcated by a yellow box in Fig.~\ref{Turbulence}(b).
We introduce two quantities for each defect. A positional error $\Delta {\bf r}$ is defined as $\Delta {\bf r}={\bf r}^\text{MED}-{\bf r}^\text{GT}$, with ${\bf r}^\text{MED}$ being the defect position vector predicted by MED, and ${\bf r}^\text{GT}$ being the ground truth position vector of the defect. 
\textcolor{black}{In Fig.~\ref{Turbulence}c, we present separate histograms for the interior and boundary regions, showing the normalized magnitude of the vector $\Delta l=|\Delta {\bf r}|/h$, where $h$ denotes the mesh's unit size in our numerical simulation.} For defect orientation, we introduce orientation angle error $\Delta \psi$ defined as the angle between the MED predicted defect orientation vector(s) and the ground truth vector(s) (Fig.~\ref{Turbulence}d). The statistics of the $\Delta \psi$ are again performed for interior and boundary regions separately. Although the boundary region exhibited marginally inferior performance in predicting defect positions, the majority of the positional deviations remain below \textcolor{black}{$2h$ (i.e., the length of one director)}. In terms of orientation, the transformer's performance was uniform across both regions, with the majority of the deviations falling below $5^\circ$. Large deviations were generally associated with defects experiencing considerable deformation or during annihilation.

We also perform error analysis on tissue images from Ref.~\cite{saw2017topological}, detailed in Appendix~\ref{SIMovie2}. Due to the absence of experimental ground truth, we calculated the scalar order parameter $S$ from the refined MED-extracted director field as benchmark. As shown in SI Movie 2, the prediction accuracy exceeded 90\% for both the number of defects and the winding number. Furthermore, the majority of the predicted defect positions deviated by a magnitude significantly less than one unit length.

To assess the computational efficiency of MED, we contrast physics-based approaches with MED, as 
\textcolor{black}{depicted} in Table~\ref{Efficiency}. For Case A pertaining to the MD simulation of Gay--Berne particles, we adopt the physics-based methodology described in Ref.~\cite{huang2023structures}. 
Case B focuses on 8 frames of active turbulence, as showcased in SI Movie 1. The computational technique written in C language is derived from Ref.~\cite{zhang2022logic}. The results of the active turbulence simulation are transferred to the physics-based method in the data form and to the MED in the image form. It should be noted that the computational time for the MED encompasses both the prediction phase and the rendering of the result image via matplotlib. In contrast, the computational time reported for traditional physics-based methods solely accounts for the numerical calculations. Analysis of Table~\ref{Efficiency} reveals that MED 
\textcolor{black}{consistently more efficient than} traditional physics-based calculations. 

\subsection{\label{Physics Information}Physics Learnt by MED}

\begin{figure}[h]
\includegraphics[width=0.49\textwidth]{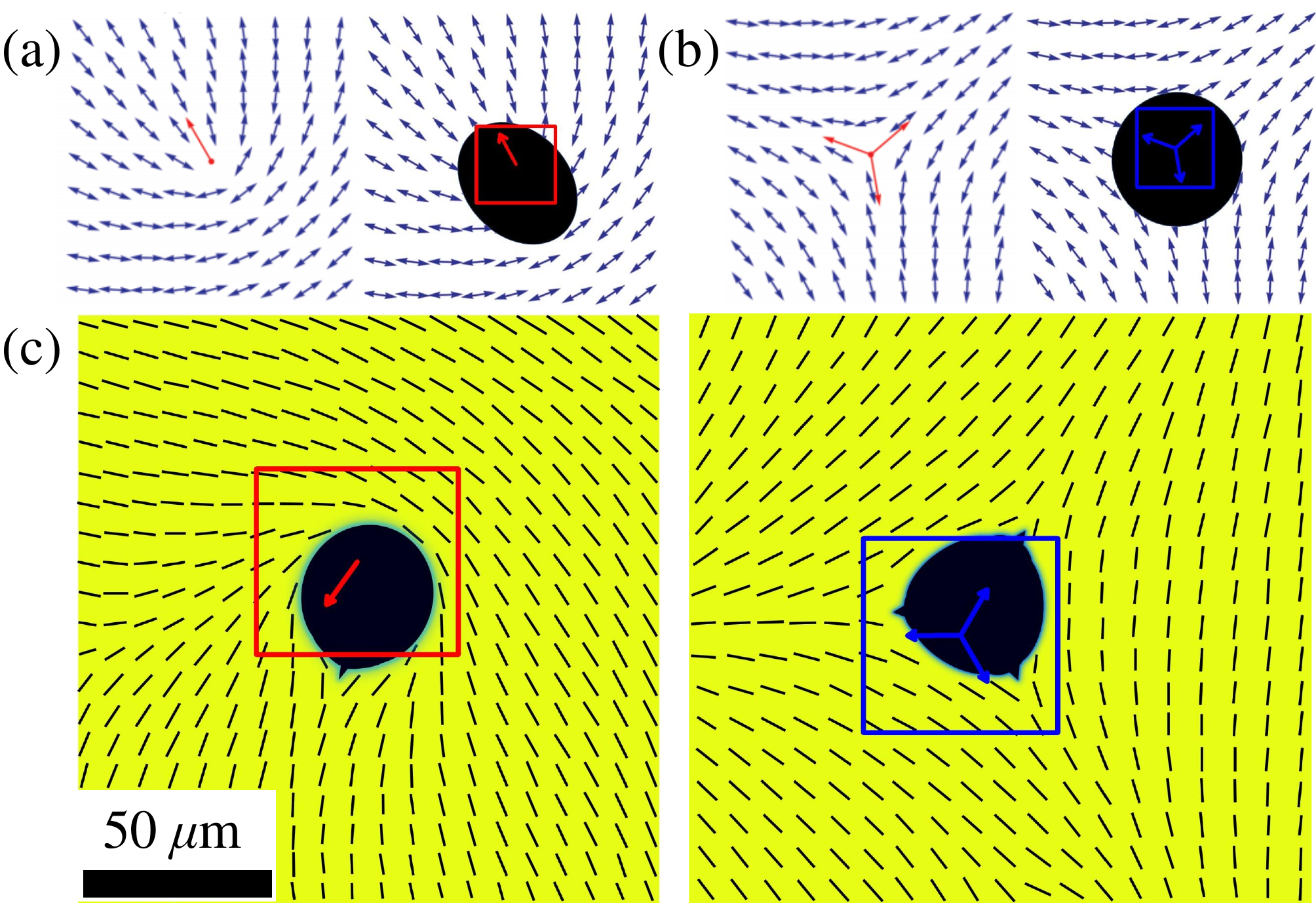}
\caption{\label{Holes} \textbf{Application of MED to $\pm1/2$ defects with giant cores.} (a) A sparse double-arrow field for a point $+1/2$ defect (left) and a $+1/2$ defect with a giant core (right). (b) A sparse double-arrow field for a point $-1/2$ defect (left) and a $-1/2$ defect with a giant core (right). (c) Experimental images of a $+1/2$ defect (left) and a $-1/2$ defect (right) \textcolor{black}{in} a lyotropic chromonic liquid crystal (disodium cromoglycate) \textcolor{black}{adapted} from Ref.~\cite{kim2013morphogenesis}. Scale bar: $50~\upmu$m.}
\end{figure}

We subsequently subjected the MED to more 
\textcolor{black}{challenging} testing scenarios that are not included in the training data. Take a sparse double-arrow vector field \textcolor{black}{(in comparison to the training data comprising simple lines)} as an example, MED can correctly identify $\pm 1/2$ defects without a visible defect core or with a giant, circular or elliptical core (Fig.~\ref{Holes}(a), (b)). \textcolor{black}{Although the double-arrow and simple-line representations of the director field exhibit the same symmetry, they contain different image details that challenge ML's interpretation. The robust performance of MED in recognizing these defect images demonstrates its generalization ability to recognize other abstract defect images.}

\begin{figure*}[htbp]
\includegraphics[width=1\textwidth]{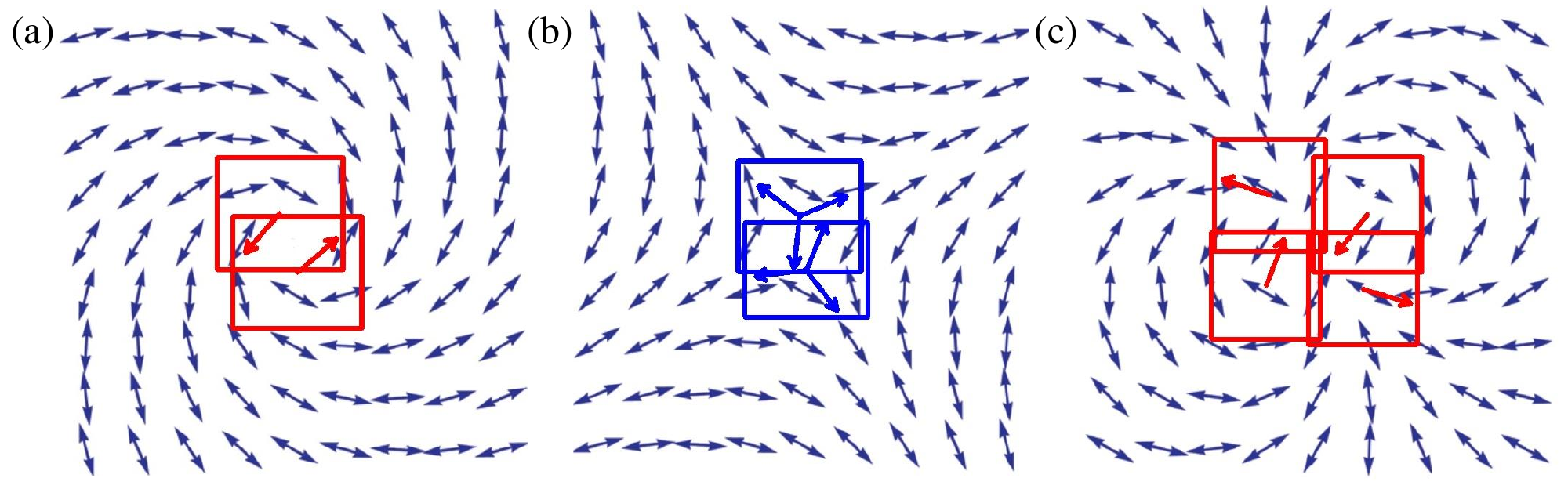}
\caption{\label{Extrapolation} \textbf{Higher charge defects not included in the training data.} (a) MED recognizes the $+1$ defect as two $+1/2$ defects. (b) A $-1$ defect identified as two $-1/2$ defects. (c) A $+2$ defect recognized as four $+1/2$ defects. These images are \textcolor{black}{adapted} from Ref.~\cite{tang2017orientation}.}
\end{figure*}

\begin{figure*}[htbp]
\includegraphics[width=1\textwidth]{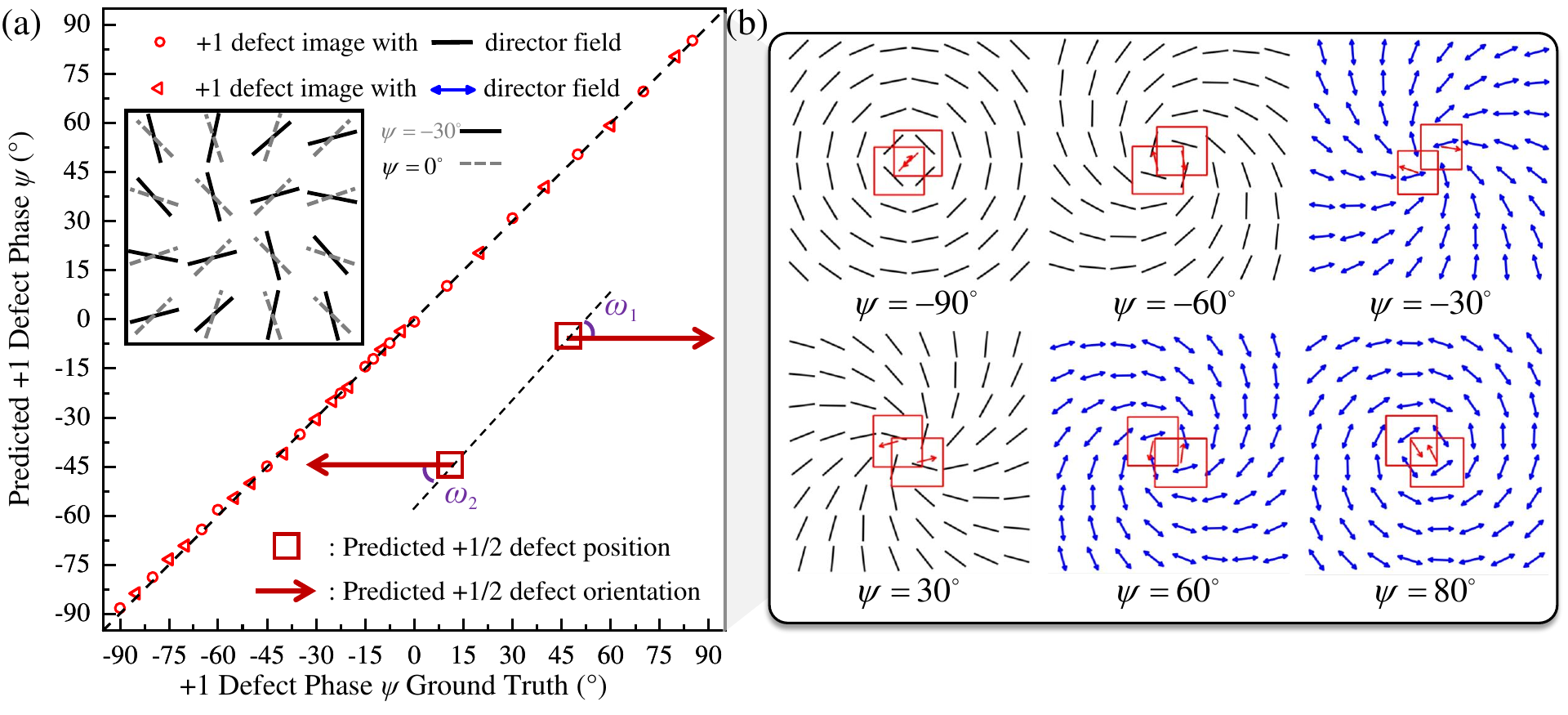}
\caption{\label{Extrapolation+1} \textbf{Phase angle prediction for $+1$ defects.} (a) Both the predicted and ground-truth values corresponding to the cases featured in SI Movie 3, supplemented by a schematic elucidating the concepts of defect phase, ${\omega _1}$, and ${\omega _2}$. (b) Some frames of the predicted $+1$ defects.}
\end{figure*}

We further apply MED to an experimental system consisting of a biphasic lyotropic chromonic LC of disodium cromoglycate solution~\cite{kim2013morphogenesis}.
During temperature change, $\pm 1/2$ defects with giant cores of specific shapes are observed~\cite{kim2013morphogenesis}.
\textcolor{black}{MED is able to identify defects in these cases too} by identifying defect locations and orientations, despite minor errors in defect positions (Fig.~\ref{Holes}(c), (d)). Note that this scenario is akin to the void defects found in microtubule-based active nematic systems~\cite{PInematic2023}, on which MED has proven its efficacy.

Considering the fact that the training data is solely composed of simple images of $\pm 1/2$ defects, we extended the prediction scope to defects with higher winding numbers (charges), such as $\pm 1$ and $+2$. The results of this extrapolation analysis are 
\textcolor{black}{depicted} in Fig.~\ref{Extrapolation}. Interestingly, MED recognizes these stranger defects as a cluster of multiple half-integer defects in the vicinity of the defect core. The total topological charge of these MED-comprehended defects coincides with the charge of the actual defect (Fig.~\ref{Extrapolation}). The ``center of mass'' of these defects appears to correspond to the location of the defect core as well (Fig.~\ref{Extrapolation}).

\begin{figure*}[htbp]
\includegraphics[width=1\textwidth]{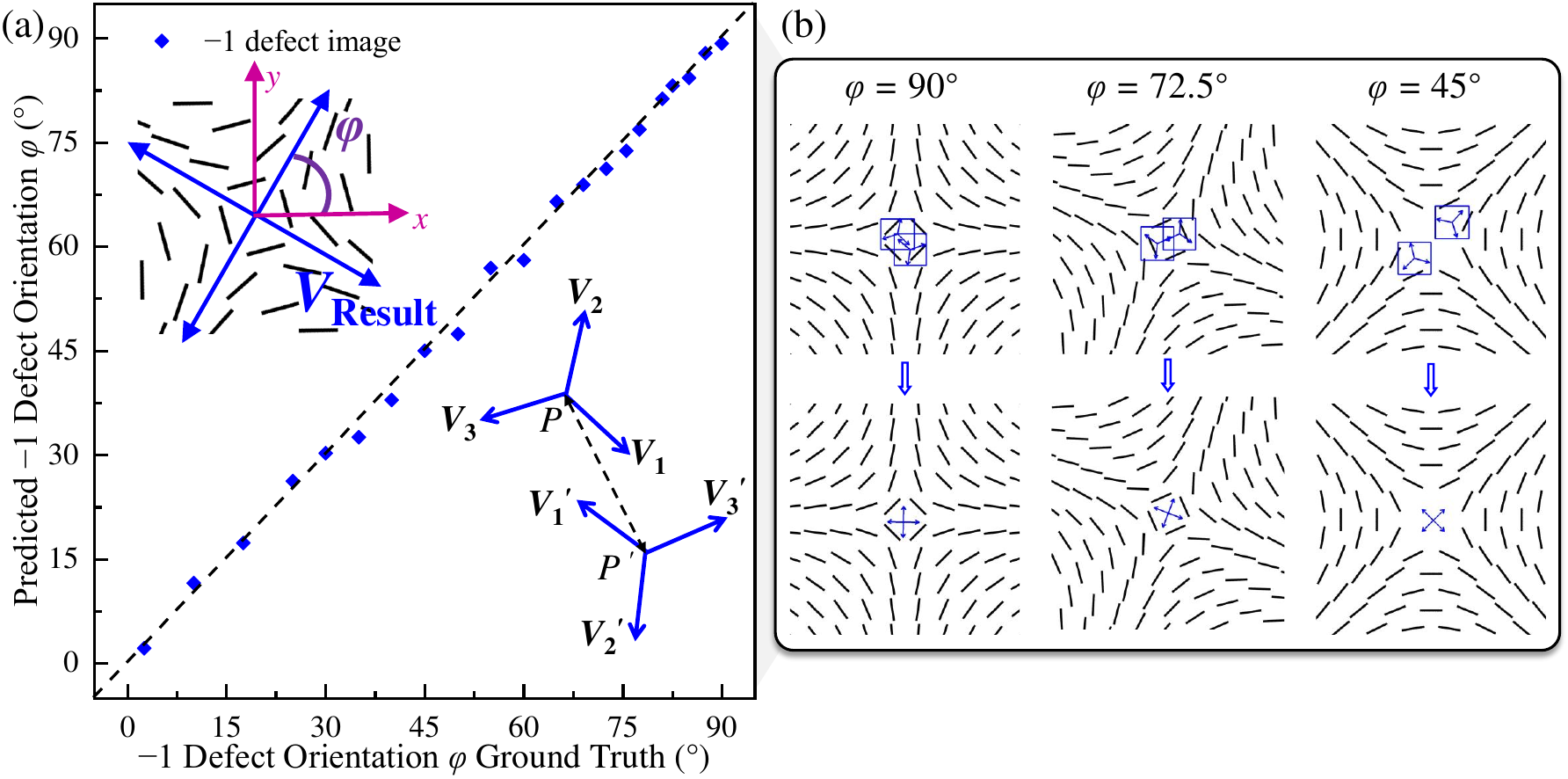}
\caption{\label{Extrapolation-1} \textbf{Defect orientation prediction for $-1$ defects.} (a) Comparison of predicted and ground-truth defect orientations for scenarios illustrated in SI Movie 4, with angles and vectors illustrated in the inset. (b) Some frames showcasing the results of the predicted $-1$ defect orientation. \textcolor{black}{The vector set ${\bf V}_{\text{Result}}$ comprises the four branches of the $-1$ defect, and the defect orientation $\varphi$ denotes how much the $x$-axis rotates counterclockwise to overlap with one branch of the defect for the first time.}}
\end{figure*}

Next, we examine whether these wrongly comprehended information by MED \textcolor{black}{(i.e., recognizing these integer defects as several half-integer defects)} contain correct structural information of the stranger defects. We first generate a sequence of images of $+1$ defects with different phase angle $\psi$ \textcolor{black}{(Using Eq.~\ref{A1} by choosing $k = 1$)}, and the prediction results are shown in Fig.~\ref{Extrapolation+1}. By introducing two angles $\omega_1$ and $\omega_2$ corresponding to the angle between the respective $+1/2$ defect and the line connecting the two defect cores, we can extract the phase angle of the $+1/2$ defect through a simple relation
\begin{equation}
\label{+1eq}
\psi = - \frac{1}{2}\left( {\frac{{{\omega _1} + {\omega _2}}}{2}} \right).
\end{equation}
We find that there is an excellent agreement between $\psi$ calculated from Eq.~\ref{+1eq} and the ground truth $\psi$, (Fig.~\ref{Extrapolation+1}(a), SI Movie 3). In Fig.~\ref{Extrapolation+1}(b), we show some $+1$ defect images overlaid with MED predictions. In these images, the director is represented by either straight lines or double arrows, exemplifying again the robust predictive capability of MED across diverse image types (SI Movie 3). 



Analogous to the generation of $+1$ defects, a sequence of $-1$ defect images was constructed using Eq.~\ref{A1}. The orientation vector of the $-1$ defect can be inferred by the orientation vectors of the two MED predicted $-1/2$ defects. We denote the two defect cores by point P and P'. Three unit vectors corresponding to the triad shape of the $-1/2$ defect orientation are denoted by ${{\bf V}}_i$ and ${{\bf V}}'_i$ for defect $P$ and $P'$, respectively, with $i=1,2,3$. The 3 vectors for each defect are ordered in counterclockwise manner, and for the same subscript $i$, ${{\bf V}}_i$ and ${{\bf V}}'_i$ are chosen to be approximately antiparallel to each other. By choosing a pair of ${{\bf V}}_i$ and ${{\bf V}}'_i$ which make the least angle with $PP'$, we can infer the vector of one of the branches of the $-1$ defect by ${\bf V}_{\text{Result}}=({{\bf V}}'_i-{{\bf V}}_i)e^{\pm j\frac{\pi}{8}(\text{sgn}({{\bf V}}_i \cdot \overrightarrow{PP'})+1)}$, where $j$ represents the imaginary unit.


The validation between the predicted orientation vector ${\bf V}_{\text{Result}}$ and the ground truth yielded a high degree of agreement, as evidenced in Fig.~\ref{Extrapolation-1} and SI Movie 4, again demonstrating that MED has comprehended the structural information of the defects.

\section{Discussion}
In this research, we present the Machine Eye for Defects (MED) system, a machine learning-based solution tailored to automatically detect and characterize defects within nematic materials with diverse textures. The MED functions analogously to the human eye, first 'visualizing' the nematic system in the form of a director field, followed by a rapid and precise analysis within this specific context. We have ingeniously integrated state-of-the-art algorithms--including Nanodet-Plus, Segment Anything Model, and Vision Transformer, with advanced computer vision techniques, allowing the MED to undertake a thorough analysis of texture images of nematic systems.
The proposed system demonstrates exceptional adaptability, precision, and efficiency in evaluating a wide range of nematic systems.
For example, MED can correctly identify defects from images of discrete \textcolor{black}{contours} as well as tight texture, and from nematic images \textcolor{black}{complicated} by void tactoids, variant lightness, or void and giant defect cores. For sparse vector fields, MED works well with double-arrow director fields.
Remarkably, MED can even comprehend higher-charge defects that 
\textcolor{black}{were not trained}. We find that MED interprets these integer defects as a cluster of 
half-integer defects, resembling human cognition pattern that human tends to comprehend unknown objects using a combination of understood concepts. Structural information about these stranger defects, including phase angle and orientation, can be correctly extracted from the positions and orientations of those half-integer defects, implying that MED has learnt the correct physics of defects.

The MED offers substantial promise for aiding the inspection of defects across various nematic systems, providing quick and automated image analysis, contingent upon an appropriate set of training data. Its practical applications transcend liquid crystals, including any ordered systems that harbor defects. Moreover, the fine-tuning of the Vision Transformer within MED might facilitate the extraction of additional essential parameters, such as elastic constants and activity \textcolor{black}{(the quantification of microscopic energy sources introduced in active nematics)}.

Despite the promising results showcased by MED, there remain several facets that could benefit from further refinement. At present, the training data solely comprise basic images of $\pm1/2$ defects. The performance of MED can be further improved if more types of defects are provided to MED. The fact that MED can comprehend stranger defects needs further studies to design better strategy for neural networks to learn physics more explicitly. MED also has the potential to learn the viscoelastic properties of nematic materials, as recent works are pursuing~\cite{colen2021machine,PInematic2023}. In \textcolor{black}{conclusion}, we expect that our work will 
\textcolor{black}{stimulate} more research efforts in this field and help extend the application of machine learning in materials science.


\begin{acknowledgments}
The authors express their gratitude to Prof. Linda S. Hirst for her insightful suggestions on predicting cytoskeletal filament images and providing experimental microtubule data. Appreciation is also due to Prof. Robin Selinger, Prof. Ivan Smalyukh, Prof. Jonathan Selinger, Prof. Jeff Z. Y. Chen and Prof. Seth Fraden for their fruitful discussions. Further thanks are extended to Mr. Wang Sheng for his valuable insights on the selection of the object detection algorithm and Transformer model. The authors would also like to acknowledge RangiLyu for the open-source contribution of the Nanodet-Plus code \cite{=nanodet} and to Alexander Kirillov et al. for making the Segment Anything model available \cite{kirillov2023segment}. This work was generously supported by the Research Grants Council of Hong Kong.
\end{acknowledgments}

\appendix

\section{\label{Technical detail}Technical Details of MED}
\subsection{Details for Preprocessing Module}
\subsubsection{\label{SAM detail}Optimization of SAM Segmentation Outcomes}
\begin{figure}[h]
\includegraphics[width=0.35\textwidth]{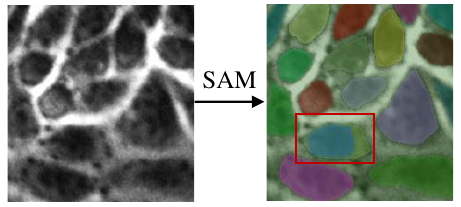}
\caption{\label{SAMoverlap} \textbf{Direct Application of SAM to a Tissue Image.} \textcolor{black}{The red box highlights the model's repetitive identification of a singular cellular structure in the epithelial cell image.}}
\end{figure}

The selection of the SAM is predicated on its zero-shot learning capability, which enables the segmentation of images with distinct contours without requiring prior knowledge. However, this generalizability comes at the expense of specificity; SAM is not innately equipped to process epithelial cell images. The direct application of SAM yields results exemplified in Fig.~\ref{SAMoverlap}. Notably, the red box indicates the model's tendency to repeatedly identify the same cellular structure, which can adversely impact the subsequent interpolation of the director field.

To mitigate this issue, we introduce a post-processing step involving the construction of a list of binary masks generated by SAM. A parameter known as Intersection over Union (IoU) is subsequently employed to evaluate and eliminate incorrectly segmented outcomes. The IoU between two masks, referred to as mask1 and mask2, is quantified as the ratio of their intersection to their union. An iterative function traverses each mask in the list (mask1) and compares its IoU with every other mask (mask2). Should the IoU exceed a predefined threshold and if mask1 possesses a smaller 'area' compared to mask2, mask1 is consequently discarded. Furthermore, in the presence of numerous spurious contours erroneously classified as cellular structures, an additional filtering mechanism based on contour size is incorporated to enhance segmentation accuracy. This approach effectively minimizes redundant identification of the same cellular structures, as illustrated in Fig.~\ref{Cells}(c).

\begin{figure*}[htbp]
\includegraphics[width=1\textwidth]{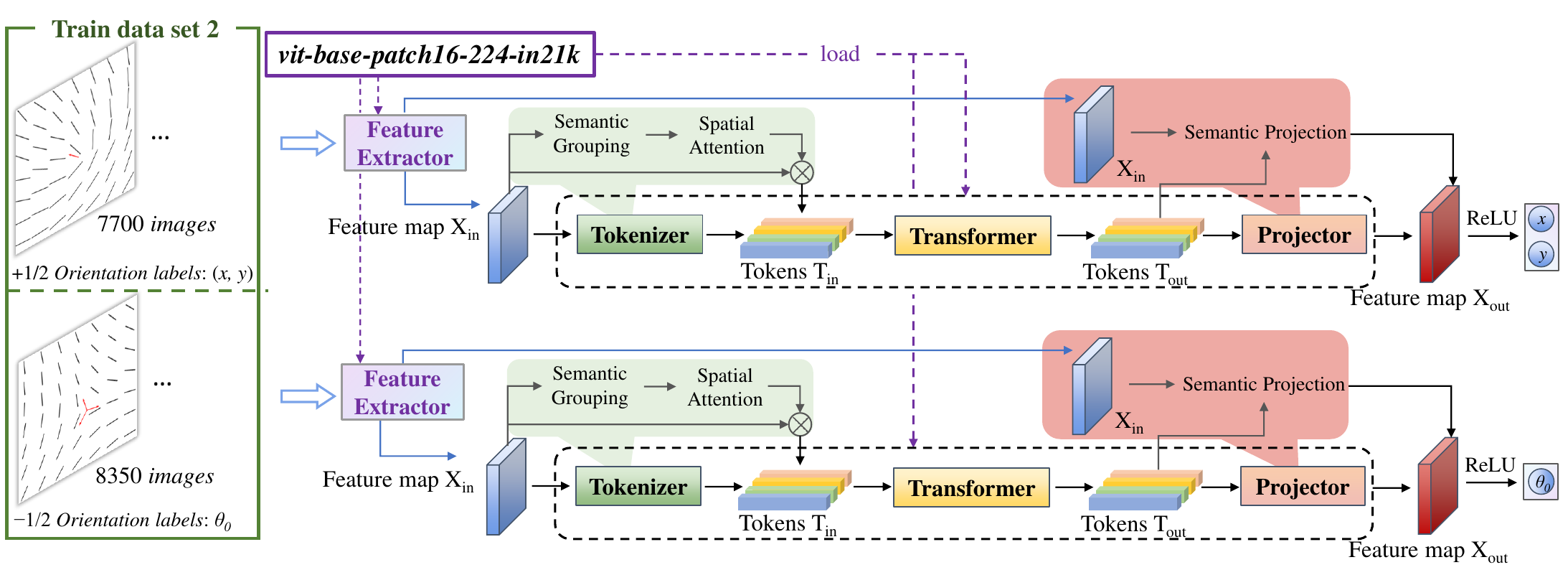}
\caption{\label{OT composition} \textbf{Structural compositions of Orientation Transformer}}
\end{figure*}

\subsubsection{\label{interpolate detail}Interpolation and smoothing method for generating the director field}
The orientation angle $\theta$, extracted from distinct contour and localization window configurations, constitutes the foundational framework for the interpolation and subsequent refinement of the director field. The desired length of the director is directly determined by the dimensions of the underlying 2D grid. Interpolation of the $\cos(2\theta)$ and $\sin(2\theta)$ components is executed via linear interpolation techniques and subsequently smoothed using a Gaussian filter, yielding the smoothed fields $\text{grid\_s2t}$ and $\text{grid\_c2t}$. The final orientation field, $\text{grid\_theta}$, over the 2D grid is computationally derived as:

\begin{equation}
\begin{aligned}
    &\text{grid\_t} = 0.5 \times \arctan2(\text{grid\_s2t}, \text{grid\_c2t}), \\
    &\text{grid\_theta} = \mod(\text{grid\_t}, \pi).
\end{aligned}
\end{equation}

Additionally, we also introduce an auxiliary method based on image super-resolution techniques to augment the textural features of actin-based images, thereby further attenuating the adverse effects of noise and blurriness.

\subsection{\label{Detectdetail}Details for Defect Detection Network}
Object detection algorithms bifurcate into Anchor-based~\cite{Fast-R-CNN,yolov2,YOLOv5} and Keypoint-based (or anchor-free)~\cite{=nanodet,CenterNet,CornerNet} classifications. Anchor-based detectors often struggle with hyperparameters, such as the numbers, sizes, and aspect ratios of anchors, which are highly dependent on the dataset~\cite{arani2022comprehensive}. Conversely, Keypoint-based strategies simplify this task by detecting objects through central keypoints, thereby avoiding the need for complex bounding box configurations. In the context of defect detection, Keypoint-based algorithms commence by identifying the core of the defect, which is then followed by a keypoint-assisted extrapolation of the defect region's size. It can offer some improvement in dealing with hyperparameter tuning  that are prevalent in Anchor-based methods.

Within the spectrum of Keypoint-based algorithms, Nanodet-Plus~\cite{=nanodet} is noteworthy for its single-stage, fully convolutional framework that judiciously balances computational efficiency and detection accuracy. This assertion is corroborated by a thorough evaluation from Arani et al.~\cite{arani2022comprehensive}, which attests to Nanodet-Plus's superior performance across multiple metrics, including accuracy, robustness to perturbations, and energy efficiency.

Object Detection Networks such as Nanodet-Plus are usually composed of three principal parts: Backbone, Neck, and Prediction heads. The Backbone, primarily responsible for feature extraction, leverages the ShuffleNetV2 algorithm \cite{ma2018shufflenet}. Serving as an intermediary, the Neck enhances the extracted features by aggregating them across multiple scales through the Path Aggregation Network for Feature Pyramid (PAFPN) \cite{liu2018path}. This is particularly salient for scenarios requiring object detection at varying resolutions. Finally, the Prediction Heads employ these refined features to generate robust and precise object classifications, thereby fulfilling the object detection pipeline. The efficacy of our defect detection network, particularly when trained on a limited dataset comprising 600 labeled images, is significantly bolstered by the Label Assignment Distillation-based training aid module integrated into Nanodet-Plus\cite{nguyen2022improving}.

\subsection{\label{detailtransformer}Details for the Orientation Transformer}

Fig.~\ref{OT composition} illustrates the structural makeup of the Orientation Transformer. After generating the 'Train data set 2' using the method detailed in Appendix~\ref{Train data}, we first invoke the Feature Extractor in the pre-training model `vit-base-patch16-224-in21k'~\cite{wu2020visual,deng2009imagenet} to convert the images in the training set into representative tensors. Simultaneously, we transform the Orientation labels from textual data into the numerical format. This process converts the training set into a Serialized Dictionary, significantly reducing the training time and enhancing the training efficacy.

The architecture we employed for both Plus Transformer and Minus Transformer mirrors the pre-training model `vit-base-patch16-224-in21k'. Prior to training, the parameters of the pre-training model are loaded, endowing the system with substantial feature extraction capabilities from the outset. Subsequently, additional training based on the data in the training set substantially enhances ViT's aptitude for extracting specific types of images we use. The Feature map, extracted by the Feature Extractor, serves as the input for the Transformer, denoted as $X_{in}$.

The Tokenizer groups pixels into semantic concepts to generate a concise set of visual tokens. The Spatial Attention mechanism dynamically allocates computational resources by focusing on significant regions rather than uniformly processing all pixels. Semantic Grouping clusters pixels into a limited number of visual tokens, each signifying a semantic concept within the image. Post tokenization, transformers model interactions between these visual tokens. Subsequently, the Projector maps these visual tokens back to pixel space to derive an augmented feature map $X_{out}$. A more detailed exploration of this series of processes can be found in Ref.~\cite{wu2020visual}. Following their passage through the ReLU activation function, the output neurons can deliver the necessary orientation results.

\section{\label{Train data}Generation method of training dataset}


Our training data are sourced from two primary methodologies: theoretical nematic field based on the Frank--Oseen theory and the hybrid lattice Boltzmann method (LBM) simulations. Specifically, `Train data set 1', used for the defect detection network, is generated by LBM simulations and the second strategy outlined in the Frank--Oseen theory. `Train data set 2', utilized for the orientation transformer, is generated by the first and third strategy in Frank--Oseen theory. Additionally, LBM simulations are also performed to produce movies depicting active turbulence movie.

\subsection{Theoretical nematic field}
\textcolor{black}{We generate three types of director field images based on the Frank--Oseen elasticity theory:}
\begin{enumerate}
    \item \textcolor{black}{\textbf{Single half-integer defect images:} This type of image involves a single, isolated half-integer defect, either $+1/2$ or $-1/2$, on a square lattice. We randomize the positions and orientations of these defects as training data.}
    \item \textcolor{black}{\textbf{Dual half-integer defect images:} This type of image involves a pair of $+1/2$ and $-1/2$ defects. The emphasis here is to include the information of defect--defect interactions.}
    \item \textcolor{black}{\textbf{Defect images using disparate elastic constants:} To enhance the model’s generalizability, this type of image focuses on a single, isolated half-integer defect with different the ratios of elastic constants.}
\end{enumerate}


\subsubsection{\label{FOone}Single Half-Integer Defect Images}
For this 
\textcolor{black}{type of image} we adopt one-elastic-constant approximation. By establishing a polar coordinate system $(r,\phi)$ centered at a defect core, the local orientation angle $\theta$ of all the directors in the lattice are determined by~\cite{tang2017orientation}:
\begin{equation}
\theta(r,\phi) = k\phi + \psi,
\label{A1}
\end{equation}
where $k$ and $\psi$ are the winding number and phase angle of the defect, respectively.

The above Equation~\ref{A1} utilizes $\phi$ to represent the angle determined by the x and y distances between the defect core and lattice points, where $\phi = \arctan (y/x)$. Here, $k$ denotes the defect charge and $\theta_0$ is the defect phase that represents an arbitrary overall rotation of the director about the z-axis. The defect orientation is extrapolated from $\psi$, described as:
\begin{itemize}[leftmargin=*]
\item For the single branch of the $+1/2$ defect orientation:
\end{itemize}
\begin{equation}
{\bf P} = \left[\cos\left(2 \psi\right), \sin\left(2 \psi\right) \right]
\end{equation}
\begin{itemize}[leftmargin=*]
\item For the three branches of the $-1/2$ defect orientation:
\end{itemize}
\begin{equation}
\begin{aligned}
{\bf V}_1 &= \left[\cos\left(\frac{2}{3} \psi\right), \sin\left(\frac{2}{3} \psi\right) \right], \\
{\bf V}_2 &= \left[\cos\left(\frac{2}{3} (\psi + \pi)\right), \sin\left(\frac{2}{3} (\psi + \pi)\right) \right], \\
{\bf V}_3 &= \left[\cos\left(\frac{2}{3} (\psi + 2\pi)\right), \sin\left(\frac{2}{3} (\psi + 2\pi)\right) \right].
\end{aligned}
\end{equation}
Incorporating Gaussian white noise into the director facilitates the generation of relevant training data for the Orientation Transformer. This training set, denoted as 'Train Data Set 2', comprises 7,700 $+1/2$ defect images and 8,350 $-1/2$ defect images.

\subsubsection{\label{FO2}Dual Half-Integer Defect Images}
For 
\textcolor{black}{this type of image}, the associated director field is derived from~\cite{tang2017orientation}:
\begin{equation}
\begin{aligned}
&\theta(\mathbf{r})= k_1 \tan^{-1}\left(\frac{y-y_1}{x-x_1}\right)+k_2 \tan^{-1}\left(\frac{y-y_2}{x-x_2}\right)  \\
& + \frac{\delta \theta}{2}\left[1+\frac{\log \left(\left|\mathbf{r}-\mathbf{R}_1\right|^2\right)-\log \left(\left|\mathbf{r}-\mathbf{R}_2\right|^2\right)}{\log \left(\left|\mathbf{R}_1-\mathbf{R}_2\right|^2\right)-\log \left(r_{\mathrm{c}}{ }^2\right)}\right]+\Theta
\end{aligned}
\label{A4}
\end{equation}

Equation~\ref{A4} describes arbitrary defect charges $k_1$ and $k_2$ at positions $\mathbf{R}_1=\left(x_1, y_1\right)$ and $\mathbf{R}_2=\left(x_2, y_2\right)$. The term $r{_\mathrm{c}}$ is indicative of the core radius. The quantities $\delta \theta$ and $\Theta$ are given by:

\begin{equation}
\begin{aligned}
\delta \theta & =\theta_2-\theta_1+k_2 \tan^{-1}\left(\frac{y_1-y_2}{x_1-x_2}\right)-k_1 \tan^{-1}\left(\frac{y_2-y_1}{x_2-x_1}\right), \\
\Theta & =\theta_1-k_2 \tan^{-1}\left(\frac{y_1-y_2}{x_1-x_2}\right) .
\end{aligned}
\end{equation}

We generated approximately 100 images, each containing both $+1/2$ and $-1/2$ defects, following the designated strategy. These images were subsequently incorporated into 'Training Dataset 1' for Nanodet-Plus training. Every image was manually annotated with meticulous care to delineate the exact nature and location of the defect present within the image. The labeling of the dataset was executed using LabelImg~\cite{LabelImg}. Given the non-uniformity of director fields in practical applications, the hybrid LBM and the Beris--Edwards equation were employed to ensure consistent alignment on the grid, as detailed in section~\ref{LBM}.

\subsubsection{Different Elastic Constants Defect Images}

Derived from the Frank elastic free energy, we have:
\begin{equation}
f=\frac{1}{2} K_{11}(\nabla \cdot \mathbf{n})^2+\frac{1}{2} K_{22}(\mathbf{n} \cdot(\nabla \times \mathbf{n}))^2+\frac{1}{2} K_{33}(\mathbf{n} \times(\nabla \times \mathbf{n}))^2
\end{equation}

For 2D considerations, only $K_{11}$ and $K_{33}$ serve as significant elastic constants, represented by the ratio $r = K_{33}/K_{11}$. The director $\bf{n}=(\cos\theta,\sin\theta)$ is represented by an angle $\theta$ in the polar coordinate $(\rho,\phi)$ centered at the defect core. \textcolor{black}{By introducing a function $f(\theta)$ as defined in below}~\cite{zhang2018interplay}:
\begin{itemize}[leftmargin=*]
\item For a $+1/2$ defect:
\end{itemize}
\begin{equation}
f(\theta) = \sqrt{\frac{1 + r + (1-r)\cos(2\theta)}{1 + r - c_0 + (1-r)\cos(2\theta)}};
\end{equation}

\begin{itemize}[leftmargin=*]
\item For a $-1/2$ defect:
\end{itemize}
\begin{equation}
f(\theta) = \sqrt{\frac{1 + r + (1-r)\cos(2\theta)}{1 + r + c_0 + (1-r)\cos(2\theta)}}.
\end{equation}
\textcolor{black}{The angle $\theta$ can be solved by}
$$
\frac{\mathrm{d} \phi}{\mathrm{d} \theta}=f(\theta).
$$
In the equation corresponding to the $+1/2$ defect, \(c_0\) is initialized as \(0.5 \times \min (1, r) \times 2\), and subsequently algorithmically modified to ensure that the numerical integration of \(f(\theta)\) over the interval \([0, \pi / 2]\) is precisely equal to \(\pi\). In contrast, for the \(c_0\) related to the $-1/2$ defect equation, the initialization is performed as \(0.5 \times \max (1, r) \times 4\), and further adjustments are made programmatically to satisfy the condition that the numerical integration of \(f(\theta)\) over the range \(\left[0, \frac{3 \pi}{2}\right]\) equals \(\pi\). The root-finding procedure detailed above ensures a consistent integration of the defect energy across diverse configurations.

We have generated a total of 2,100 images of $+1/2$ defects by systematically varying both the value of $r$ and the orientations of the defects. This variation led to the manifestation of distinct defect profiles, such as U-shaped or V-shaped $+1/2$ defects, consistent with the observations detailed in Ref.~\cite{zhang2018interplay}. In a parallel effort, we produced 1700 images of $-1/2$ defects, once again modulating the value of $r$ and the orientations, thereby allowing for an extensive investigation into the full spectrum of conceivable defect structures. These images have also been incorporated into `Train Data Set 2'.

\subsection{\label{LBM}Hybrid LBM method}
The 2D nematic considered here can be described by a tensorial order parameter ${\bf Q}$ and a velocity field ${\bf u}$, respectively. For a uniaxial nematic LC, ${\bf{Q}} = S({\bf{nn}} - {\bf{I}}/3)$, where the unit vector $\bf{n}$ represents the nematic director field, $S$ is the scalar order parameter of the nematic LC, and ${\bf I}$ is the identity tensor. By defining the strain rate ${\bf D}=(\nabla{\bf u}+(\nabla{\bf u})^T)/2$ and the vorticity $\boldsymbol{\Omega}=(\nabla{\bf u}-(\nabla{\bf u})^T)/2$, we introduce an advection term
${\bf{S}} = (\xi {\bf{D}} + {\boldsymbol{\Omega}}) \cdot ({\bf{Q}} + \frac{{\bf{I}}}{3}) + ({\bf{Q}} + \frac{{\bf{I}}}{3}) \cdot (\xi {\bf{D}} - {\boldsymbol{\Omega}}) -2\xi ({\bf{Q}} + \frac{{\bf{I}}}{3})({\bf{Q}}:{\bm{\nabla}} {\bf{u}})$, where $\xi$ is related to the constituent's aspect ratio. In our simulation, we consider flow-aligning liquid crystal by setting $\xi=0.8$.

The governing equation of the {\bf Q}-tensor, the Beris--Edwards equation, is \cite{beris1994thermodynamics}
\begin{equation}
 \frac{\partial {\bf{Q}} }{{\partial t}} + {\bf{u}} \cdot {\bm{\nabla}} {\bf{Q}} - {\bf{S}} = \Gamma {\bf{H}},
 \label{beris}
\end{equation}
where $\Gamma$ is associated with the rotational viscosity of the nematic liquid crystals (LCs), expressed as $\gamma_1=2S_0^2/\Gamma$~\cite{denniston2001lattice}. The molecular field ${\bf H}$ is defined by ${\bf H} = -(\frac{\delta F}{\delta {\bf Q}}- \frac{{\bf I} }{3}\tr(\frac{\delta F}{\delta {\bf Q}})) $, which drives the system towards thermodynamic equilibrium with the free energy functional $F=\int_{V} f dV $. Here, $f$ represents the free energy density in the bulk, given by~\cite{de1993physics} $f_{\text{LdG}}=\frac{A_0}{2}(1-\frac{U}{3})\tr({\bf Q}^2) - \frac{A_0 U}{3}\tr({\bf Q}^3) + \frac{A_0 U}{4} (\tr({\bf Q}^2))^2+\frac{L}{2}\left(\nabla {\bf Q}\right)^2$, where $A_0$ is a phenomenological coefficient setting the energy density scale and $U$ is a material constant which controls the magnitude of $S_0$ via $S_0=\frac{1}{4}+\frac{3}{4}\sqrt{1-\frac{8}{3U}}$. $L$ is the elastic constant. The nematic coherence length, denoted by $\xi_{\text{N}}=\sqrt{L/ A_0}$, ascertains the defect core size and functions as the essential length scale in our description of nematic materials.

Using the Einstein summation rule, the Navier--Stokes equation for active nematics can be written as:
\begin{multline}
 \rho \left(\partial _t+u_{\beta}\partial_{\beta}\right)u_{\alpha}=\partial _{\beta}\Pi_{\alpha\beta}+\eta\partial _{\beta}\left[\partial_{\alpha}u_{\beta} \right. \\
 +\left.\partial_{\beta}u_{\alpha}+(1-3\partial_{\rho}P_0)\partial_{\gamma}u_{\gamma}\delta_{\alpha\beta}\right].
 \label{NS}
\end{multline}
The stress ${\bf \Pi}$ is defined as
\begin{multline}
 \Pi_{\alpha\beta}=-P_0\delta_{\alpha\beta}-\xi H_{\alpha\gamma}(Q_{\gamma\beta}+\frac{1}{3}\delta_{\gamma\beta}) \\
 -\xi(Q_{\alpha\gamma}+\frac{1}{3}\delta_{\gamma\beta})H_{\gamma\beta} +2\xi(Q_{\alpha\beta}+\frac{1}{3}\delta_{\alpha\beta})Q_{\gamma\epsilon}H_{\gamma\epsilon}\\
 - \partial _{\beta}Q_{\gamma\epsilon}\frac{\delta F}{\delta \partial_{\alpha}Q_{\gamma\epsilon}}+Q_{\alpha\gamma}H_{\gamma\beta}-H_{\alpha\gamma}Q_{\gamma\beta}-\zeta Q_{\alpha\beta},
\end{multline}
where $\eta$ is the isotropic viscosity, while the hydrostatic pressure $P_0$ is expressed by~\cite{fukuda2005interaction} $P_0=\rho T-f$. The temperature $T$ correlates with the speed of sound $c_s$ through the relationship $T=c_s^2$. The activity parameter $\zeta$ encompasses the local stress stemming from the spatial gradients of the nematic order parameter~\cite{hatwalne2004rheology,marenduzzo2007steady}. We address the evolution equations employing the finite difference method, and the Navier--Stokes equation~\ref{NS} is tackled using a lattice Boltzmann method on a D3Q15 grid~\cite{guo2013lattice}
. The validity of our model and its implementation has been substantiated by comparing our simulation outcomes~\cite{zhang2016lattice,wang2021interplay} in both passive and active nematic systems with predictions derived from the Ericksen--Leslie--Parodi (ELP) theory~\cite{kleman2003soft,ericksen1969continuum,leslie1966some,parodi1970stress}.

\begin{figure*}[htbp]
\includegraphics[width=1\textwidth]{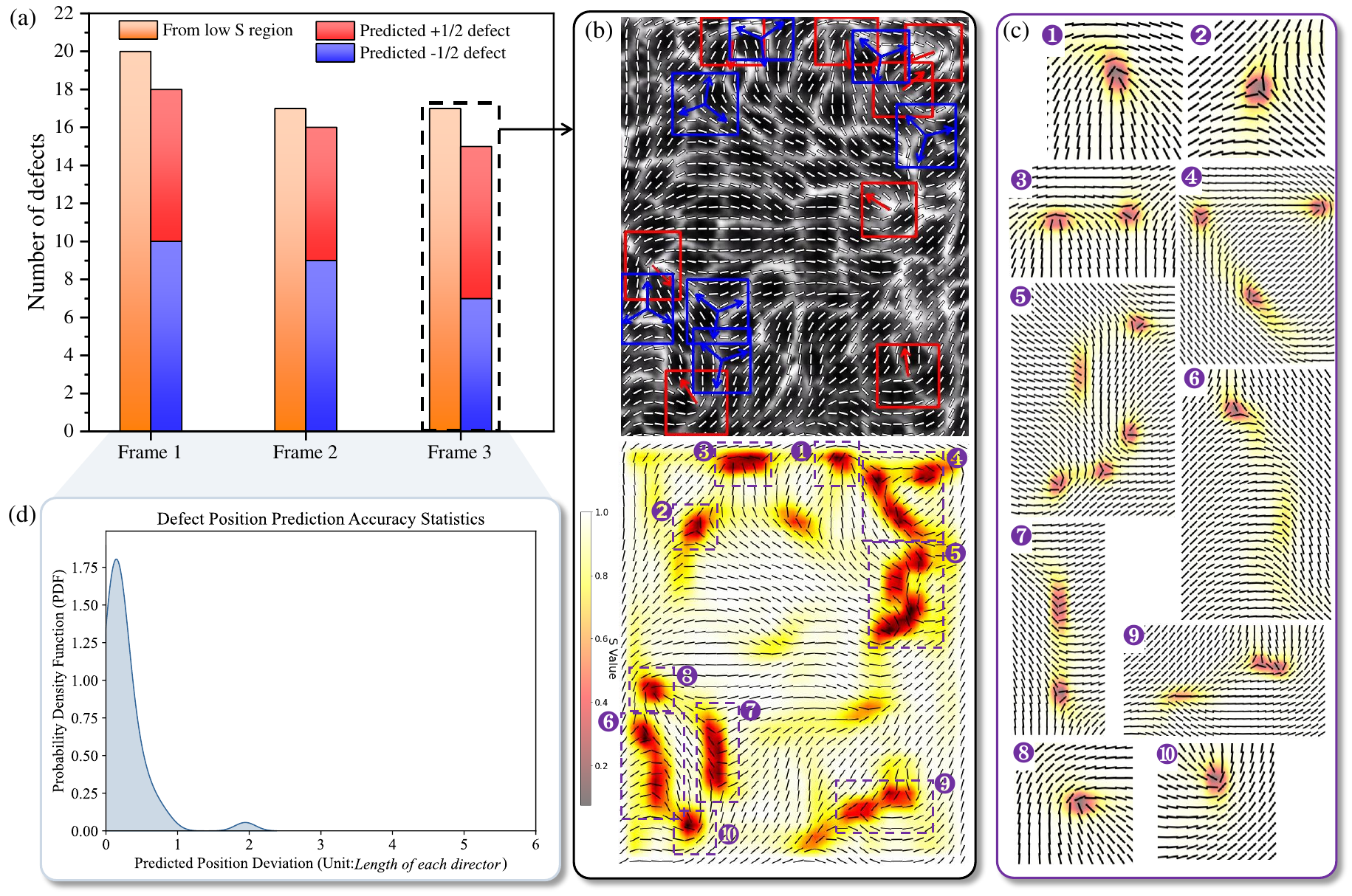}
\caption{\label{Cells Statistic} \textbf{Quantitative statistics of SI Movie 2.} (a) Comparison of defect number between MED calculations and manual counting ($S < 0.5$ regions are judged as defects). (b) Upper portion illustrates MED's prediction for Frame 3, while the lower part shows the distribution of order parameter $S$ calculated from the director field data; (c) Enhanced interpolation is performed on the 10 low-$S$ regions in the $S$ distribution image to pinpoint the locations of defect cores. (d) Histogram of deviation of MED predicted defect core positions from those manually measured, \textcolor{black}{the length of each director is close to $2~\upmu$m.}}
\end{figure*}

Under the one-elastic-constant approximation  ($L=L_1=0.1, L_2=L_3=L_4=0$), we conducted simulations for both passive and active nematics \textcolor{black}{and applied periodic boundary conditions for both cases.} For the passive nematic case, which constituted the `Train Data Set 1' for defect detection network, we embarked on a nascent defect annihilation simulation by introducing a pair of $\pm 1/2$ defects into the system. The dataset was enriched by applying rotation and mirror transformations to the extracted images from the simulation, ultimately amassing around 500 images for the training of Nanodet-Plus. The labeling procedure was analogous to that described in Appendix~\ref{FO2}. What's more, the active turbulence scenario in SI Movie 1 was utilized to assess the accuracy and efficiency of MED. Within the active turbulence simulation, an activity parameter of $\zeta = 0.007$ was used.

\section{\label{SIMovie2}Statistical details for SI Movie 2}

To rigorously assess the robustness of MED in predicting images with distinct contours, we extended our predictions to additional Epithelial Cell images from Ref.~\cite{saw2017topological}. Three images are analyzed, and the corresponding predictions are present in SI Movie 2. Given the challenge in discerning defect positions from experimental images with naked eye, we recorded the extracted director field data, which enables us to compute a map of the scalar order parameter $S$, from which we manually identify defects according to an empirical relation $S < 0.5$. Note that MED's accuracy requirements for the extracted director field are not stringent and do not rely on the $S$ director.
SI Movie 2 includes a comparative analysis of defect positions filtered by $S$ and the MED predictions. Out of three image frames, 54 defects were identified based on $S<0.5$ and experience, with MED accurately predicting 49 of them (26 $-1/2$ defects and 23 $+1/2$ defects), as demonstrated in Fig.~\ref{Cells Statistic}(a). The predicted defect charges passed individual verification with no errors, achieving a prediction accuracy of 90.74\%. Furthermore, we conducted a statistical analysis of the positional deviation of the predicted defect cores, as summarized in Fig.~\ref{Cells Statistic}(d). The data reveal that the deviation for most predicted defect positions is significantly less than the length of one director. For a given image of the director field, MED is capable of accurately ascertaining the charges, positions, and orientations of most topological defects within a time frame of approximately 3 seconds, following the extraction of the director field. In contrast, the calculation of the order parameter $S$, utilizing more precise director field data, necessitates a substantially longer time period exceeding 3 seconds. This process also involves judging whether a feature is a defect based on empirical assessment, but without providing orientation information. The described efficiency and comprehensiveness underscore the advantage of using ML methods.


\section{Data availability}
Training data set and trained model for defect detection network and Orientation transformer are available on a Zenodo repository (https://zenodo.org/records/8381966). Some sample code of MED that can predict defect positions, wingding numbers and orientations are also available on the Zenodo repository. More code are available from the corresponding author upon reasonable request. 

\nocite{*}

\bibliography{MED}

\end{document}